\documentclass[usenatbib,useAMS]{mn2e}

\usepackage{color}
\usepackage{graphicx}
\usepackage{amsmath}
\usepackage{amssymb}

\title[Population statistics of beamed sources. I: A new model for blazars]{Population statistics of beamed sources. I}
\author[Liodakis and Pavlidou]
{I.Liodakis$^{1}$\thanks{liodakis@physics.uoc.gr} and V. Pavlidou$^{1,2}$\\
$^{1}$Department of Physics and ITCP\thanks{Institute for Theoretical
  and Computational Physics, formerly Institute for Plasma Physics}, University of Crete, 71003, Heraklion, Greece\\
$^{2}$Foundation for Research and Technology - Hellas, IESL, Voutes, 7110 Heraklion, Greece\\
}
\begin{document}



\maketitle

\label{firstpage}

\begin{abstract}
Observations of blazar jets are shrouded in relativistic effects,
thus hindering our understanding of their intrinsic properties and
dominant physical processes responsible for their generation,
evolution, radiation and particle emission. In this work we focus on extracting information about timescales in the jet rest frame using a population-modeling approach. We employ Monte Carlo simulations to derive a simple population model
for the intrinsic unbeamed luminosities and the Lorentz $\Gamma$ of
blazar jets that adequately describe the observed redshift and
apparent superluminal speed distributions for flux-limited blazar samples. We derive separate models for BL Lacs and Flat Spectrum Radio Quasars. We then use these models to compute the
predicted distribution of Doppler factors in each blazar class, and
address the following questions: (a) What is the relativistically
induced spread in observed timescales (e.g., event duration, time lags) in a flux-limited sample of relativistic jets? (b) Could differences between BL Lacs and FSRQs observed in the time domain be attributed to differences in beaming between the two populations? (c) Is there a statistically preferred amount of beaming in a flux-limited sample? How large are statistical deviations from that preferred value? We use our findings to propose promising approaches in phenomenological studies of timescales in blazar jets. 
\end{abstract}


\section{Introduction}\label{intro}

Blazars are active galactic nuclei (AGN) with jets oriented within a small
angle from our line of sight \citep{Blandford1979}. Because
of their preferential alignment, their observed properties are
obscured by relativistic effects such as Doppler boosting of their
emission, compression of variability timescales, and apparently superluminal motions of resolved jet components. Small variations in the degree of alignment with the line
of sight can result to a large scatter in the resulting observable
quantities from otherwise similar sources.
 These effects complicate our understanding of their 
intrinsic properties and the processes relevant to their central engines.

 Blazars are extremely variable broadband emitters. Despite decades of
systematic study of their variability properties across the
electromagnetic spectrum, little is known
regarding the variability properties of blazars as a population {\em
  in the jet rest frame,} because of the difficulties involved in directly
measuring the Doppler factor of blazar jets. One approach that has
been used to that end is to identify the shortest-duration flare that
has been observed in an object, and compare it with some known rest-frame
timescale one can associate with the source (for example the
light-crossing time of the central black hole, see
e.g. \cite{Aharonian2007}; or, in radio wavelengths, the size of an
emission region of known brightness temperature such as the
equipartition,
e.g. \citealp{Readhead1994,Lahteenmaki1999-III,Hovatta2009}.) 
These approaches are valuable since they constitute our only way to assess Doppler boosting on a source-by-source basis. However, these methods have well-known drawbacks. First of
all, the shortest observed timescale only gives a limit to the
observed Doppler factor even if all other assumptions hold exactly, as
blazar lightcurves
have power-law power spectra \citep{Abdo2010,Chatterjee2008}. Second, the use of
other known physical parameters of the jet, such as the black hole
mass, to derive Doppler factors, prohibits any correlation studies
between these parameters and Doppler factors.

Although these difficulties cannot be easily circumvented on a
blazar-by-blazar basis, the connection between observed and 
intrinsic (jet rest frame) variability properties for blazars {\em as a
population} can be assessed in a more straight-forward way.  
Assuming randomly distributed
line-of-sight orientations for active galactic nuclei jets, we can
seek the distribution of intrinsic jet parameters (rest-frame
luminosities and Lorentz factors) that best reproduce well-defined
observables  (such as
redshifts and apparent superluminal speeds) rather than the Doppler
factors themselves.  Then, from these
distributions, we can calculate the Doppler factor distribution for the
blazar population. We can then use this distribution to deconvolve the
Doppler-factor effects on the population as a whole, and gain a statistical 
insight on the rest-frame variability properties. In this way, we can
address questions such as:
\begin{itemize}
\item[(a)] What is the relativistically induced spread in observed
  event timescales? Could blazar behaviors in the time domain observed
  to be very varied be in fact very similar in the jet rest frame? 
\item[(b)] How different is the beaming between sources in flux-limited
  samples? Lacking any additional information, is it useful to make
  statistics-based assumptions for the viewing angle of a single
  source? 
\item[(c)] Do BL Lacs and Flat Spectrum Radio Quasars (FSRQs) have different
  beaming properties? Could differences observed between them in the
  time domain be attributed to how relativistic effects differently
  affect each population? 
\end{itemize}

Our primary motivations for this work are: (i) the study of blazar events of a
well-defined duration, such as swings of the polarization angle seen
in the optopolarimetric study of blazars (especially in light of
currently ongoing large-sample, high-cadence optopolarimetric
monitoring programmes such as RoboPol, \citealp{Pavlidou2014}); and (ii) the study of
time lags between different types of events (e.g., between flares at different
wavelengths, or between polarization angle swings and flares). However
our results are general and can be applied to any time-domain
studies of blazar jets. 

This paper is organized as follows. In \S \ref{model} we describe our
model for the blazar population. In \S \ref{sample} we discuss the 
sample we use in order to derive model parameters. In \S
\ref{parameters} we present our optimization procedure: our model acceptability criteria, the set of observables we have required our model to
reproduce, and our optimization algorithm. 
In \S \ref{results} we present our results for our samples and in \S \ref{conclusions} and \S \ref{discussion} we discuss the validity of our model and computations,their implementation and the conclusions derived from this work.

The cosmology we have adopted throughout this work is $H_0=71$ ${\rm km \, s^{-1} \, Mpc^{-1}}$, $\Omega_m=0.27$ and $\Omega_\Lambda=1-\Omega_m$ \citep{Komatsu2009}. This choice was made so that our cosmological parameters agree with the MOJAVE(Monitoring Of Jets in Active galactic nuclei with VLBA Experiment, \citealp{Lister2005}) analysis \citep{Lister2009-2}.  

\section{Model}\label{model}

The purpose of our model is to simulate observations from a flux-
limited sample of blazars. The population model will consist of a joint
distribution for the intrinsic, unbeamed luminosity at a specific
(radio) wavelength at different redshifts, and for the jet Lorentz factor. 
From these, assuming a uniform distribution of viewing angles {\em for
  the population}, we can
then calculate our observables: flux density as measured by 
the observer, and redshifts and apparent superluminal speeds for
sources above a certain flux limit. 

Once we identify model parameters for which the distribution of observables are
adequately reproduced for a well-defined sample, we can then calculate
the distribution of hard-to-observe quantities {\em for the sample}, such as 
viewing angles and Doppler factors. 

The idea of building a population model for beamed sources is not
new. Models have been fitted by, e.g
\citet{Padovanni1992,Padovani1992-2,Vermeulen1994,Lister1997}. 
More recently, the observation of jet speeds by the MOJAVE program for a large,
flux-limited sample of blazars, the large majority of which have
measured redshifts, has provided an unprecedented set of observables against
which such models can be tested and re-optimized (see, e.g., \citealp{Cara2008}). 

Here, we optimize a new population model for the blazar population. The reasons why this was necessary in order to address the specific issues we are after, and the associated ways our model differs from past work using the MOJAVE dataset are summarized below.
\begin{itemize}
\item {\em We treat BL Lacs and FSRQs as distinct populations.} As the
  interpretation of optical wavelength data is one of our primary
  motivations, deriving (potentially) different models for optically
  distinct classes of blazars\citep{Giommi2012} is important in order to assess any
  differences between these classes in the time domain. 
\item {\em We do not seek to reproduce the normalization of the
    luminosity function.} As our luminosity function model is that of
  pure luminosity evolution (up to a maximum redshift), this implies
  that the total number density of sources remains constant, and the
total number of sources in a redshift interval is simply proportional
to the volume element in that same interval. Our purpose is to
determine the distribution of these sources among different (unbeamed)
luminosity values rather than the total number of sources in any
interval. For this reason, we have the flexibility to simply remove
from our sample any sources with unknown redshift. This choice does
not affect our luminosity distribution under one of the following two
assumptions: if {\em either} the sources without a redshift
measurement have the same redshift distribution as the sources
  with measured redshifts; {\em or} all of the sources without a
  redshift measurement reside at higher redshifts than the sources with
  measured redshifts, due to a bias in our ability to measure
  redshifts favoring nearby sources. In the latter case, we will
  incorrectly surmise that the luminosity distribution sharply
cuts off above a redshift that is too low, but the luminosity
  distribution at redshifts where we can measure it will be in general
  correct. 
\item {\em We do not include individual blazar flux densities in the set of
    observables the distributions of which our model has to
    reproduce.} The reason is that blazars are known to show significant variability in all wavelengths and
across timescales, making flux density an unreliable observable for model
fitting. Instead, we only use the measured and simulated blazar flux
densities to determine if a specific source would make or not the cut for
inclusion in a flux-limited sample. 
\item {\em We simultaneously fit the unbeamed luminosity and bulk
    Lorentz factor distributions.} Since we are primarily interested
  in blazar timescale modulation factors in a flux-limited sample, the
  unbeamed luminosity function and the bulk jet Lorentz factor distribution are
  of equal importance in producing the results of interest. For this
  reason, we want to avoid taking one of the two as input from past
  work and only fitting the other. 
\item{\em We focus on simplicity in both our models and our acceptability criteria.} Given that the systematic uncertainties (flux
  variability, redshift incompleteness, variation of component speeds
  within a single blazar) can be very significant, it is likely that
  as our understanding of blazar physics improves, our observables
  themselves will change, and any complex population model and/or sophisticated
  fit will change with them. For this reason, in this work we try to
  keep both the models as well as our statistical treatment as simple
  as possible, and our model acceptability criteria generous. 
\end{itemize}

\subsection{Unbeamed Luminosity Function}\label{Luminosity}

We assume that the jet Lorentz factor and the intrinsic, unbeamed
monochromatic luminosity are uncorrelated.  Expecting the blazar unbeamed luminosity function
to evolve with redshift, we have adopted a pure luminosity evolution
model, with a single power law between values $L_{\nu, \rm min}$ and
$L_{\nu, \rm max}$ of the form 
\begin{equation}\label{lumfunc}
n(L_\nu ,z)\propto \left( \frac{L_\nu }{e^{{T(z)}/\tau}} \right)^{-A}, 
\end{equation}
where $n$ is the comoving number density of blazars, L is the intrinsic luminosity, $\tau$ is the evolution parameter in units of Hubble time
\citep{Padovanni1992}, and T(z) is the look-back time at a given redshift,
\begin{equation}
T(z)=\frac{1}{H_0}\int\frac{da}{a\sqrt{\Omega_ma^{3}+\Omega_\Lambda}},
\end{equation}
where $a=(1+z)^{-1}$ is the scale factor of the Universe. We take
Eq.~(\ref{lumfunc}) to be valid up to a maximum redshift, equal to the
largest measured redshift included in our sample. At higher redshifts,
we assume that the normalization of the luminosity function sharply
declines with redshift. We do not implement a specific functional form
for this decline, simply assuming that it is steep enough so that no
higher-$z$ source makes it into our flux-limited sample. 

The probability density function (PDF) of the luminosity has the form
 \begin{equation}
p(L_\nu)=C_2\left( \frac{L_\nu}{e^{{T(z)}/\tau}} \right)^{-A},
\end{equation}
and the value of constant $C_2$ can be obtained by the requirement that
the probability density integrates to $1$, 
\begin{equation}
C_2=\frac{(-A+1)e^{{-AT(z)}/\tau}}{L_{\nu, \rm max}^{-A+1}-L_{\nu, \rm min}^{-A+1}}.
\end{equation}
The cumulative density function (CDF) of this distribution is 
\begin{equation}
{\rm CDF (L_\nu)} = \frac{L_\nu^{-A+1} -L_{\nu, \rm
    min}^{-A+1}}{L_{\nu, \rm
    max}^{-A+1}-L_{\nu, \rm min}^{-A+1}} . 
\end{equation}

In the pure luminosity evolution model, sources become brighter with
look-back time while maintaining a constant comoving number
density \citep{Padovanni1992}. 
Then, the number of sources $N$ in a redshift interval from $z$ to
$z+dz$ is proportional to the comoving volume element $dV$ ($N = n dV$), where 
\begin{equation}
dV=\frac{c}{H_0}\frac{4\pi{d_c^2}dz}{\sqrt{\Omega_m(1+z)^{3}+\Omega_\Lambda}},
\end{equation}
 and $d_c$ is the comoving distance, 
\begin{equation}
d_c = \frac{c}{H_0} \int_0^z \frac{dz'}{\sqrt{\Omega_m(1+z')^3+\Omega_\Lambda}}.
\end{equation}

\subsection{Lorentz Factor Distribution}\label{Lorentz}

We assume a single power law of the Lorentz factor $\Gamma$ of the jet
is sufficient to
describe any sample, as suggested by \cite{ Padovanni1992} and
\cite{Lister1997}. 
The probability density function p($\Gamma$) has the form of,
 \begin{equation}
p(\Gamma)=C_1\Gamma^{-\alpha},
\end{equation}
where $C_1$ is a normalization constant,
\begin{equation}
C_1=\frac{-\alpha+1}{\Gamma_{max}^{-\alpha+1}-\Gamma_{min}^{-\alpha+1}},
\end{equation}
with $\Gamma_{min}=1$ and $\Gamma_{max}\approx\beta_{app}^{max}$\citep{Vermeulen1994}.

\subsection{Viewing Angles}

We assume a random viewing angle ($\cos\theta$  uniformly distributed from 0 to 1.) Blazar jets are closely aligned with our line of
sight, and we expect this to hold for the observed flux-limited sample
as  sources with viewing angles larger than a few degrees will not have
sufficient Doppler boosting in order to pass the flux-density limit in
wavelengths where blazars are the dominant AGN class. 

\subsection{Derived Quantities}

Once the intrinsic unbeamed monochromatic luminosity $L_\nu$, the bulk Lorentz factor
$\Gamma$ of the jet, the redshift $z$, and the viewing angle $\theta$
are known for a source, then we can calculate a series of derived
quantities, including: 
\begin{itemize}
\item the apparent
superluminal speed of the jet, 
\begin{equation}
\beta_{\rm app} = \frac{\beta \sin \theta}{1-\beta \cos \theta}\,
\end{equation}
 where $\beta \lesssim 1$ is the speed of the jet
in units of the speed of light, which is connected to the Lorentz factor
through \begin{equation}
\Gamma=\frac{1}{\sqrt{1-\beta^2}}\,;
\end{equation}
\item the Doppler factor, 
\begin{equation} 
 D=\frac{1}{\Gamma(1-\beta\cos\theta)}\,;
 \end{equation} 
\item the apparent timescale $\Delta{t'}$ of events that have a
  duration $\Delta{t}$ in the jet rest frame, 
\begin{equation}
\Delta{t'}=\frac{1+z}{D}\Delta{t}\,;
\label{eqtscale}
\end{equation}
\item and the observed monochromatic flux density, 
\begin{equation}\label{flux}
S_\nu=\frac{L_\nu D^p}{4\pi{d_L^2}}(1+z)^{1+s}.
\end{equation}
In Eq.~(\ref{flux}), $L_\nu$ is the unbeamed monochromatic luminosity, 
$d_L=(1+z)d_c$ is
the luminosity distance at a given redshift, and (s) is the spectral
index, with its sign defined through $S_\nu \propto \nu^s$.
The exponent $p$ is
given by $p=2-s$ for
the continuous and $p=3-s$ for the discrete jet cases
\citep{Ghisellini1993}. Following \citealp{Lister2009-2} we adopt the continuous jet case ($p=2-s$). For FSRQs we take $s=0.37$,(a value calculated
in \citet{Lister2009-2} by fitting an envelope to the $\beta_{\rm app} $ vs 15 GHz luminosity
plot for MOJAVE sources); for BL Lacs we take $s=0$ \citep{Urry1991,Lister1997,Cohen2007}.
\end{itemize}

\section{Sample}\label{sample}

Our sample consists of sources with a measured redshift from
the MOJAVE survey statistically complete, flux-limited sample
\citep{Arshakian2006, Pushkarev2009, Pushkarev2012}. We have excluded
objects that have shown any abnormal behavior, inward motion or poor apparent velocity measurement, as indicated by
\citet{Lister2009-2,Lister2013}. We also removed objects 0805-077 and
0642+449 because they were outliers, possibly indicating unique or
abnormal properties: 0805-077 exhibits a far greater apparent speed
than any other object ($\beta_{app}\approx 50$)  and 0642+449 has an
untypically high redshift($z=3.396$).

Our final sample consists of the 74 FSRQs and 16 BL Lacs shown in table 1 and 2 respectively.

Since we will be using MOJAVE observations at 15 GHz for the component
speeds, the flux densities and luminosities we use in this work will
also refer to a frequency of 15 GHz. The flux limit of our adopted sample,
which we will also impose on our simulated data, is 1.5 Jy. 

\begin{table*}
\setlength{\tabcolsep}{11pt}
 \centering
 \begin{minipage}{140mm}
 \centering
  \caption{FSRQS sample}
  \label{tab:FSRQs}
  \begin{tabular}{@{}ccccccc@{}}
  \hline
   Object Name & $\beta_{app} $ & Redshift  &  Object Name & $\beta_{app} $ & Redshift    \\
 \hline
 0016+731  & 6.74 & 1.781  &  1219+044 & 2.35 & 0.965  \\
  0059+581 & 8.705 & 0.644  &  1222+216 & 15.882 & 0.432  \\
  0106+013 & 24.04 & 2.099   & 1226+023 & 9.643 & 0.158  \\
  0119+115 & 17.1 & 0.570   &  1334-127 & 6.99 & 0.539      \\
  0133+476 & 11.5 & 0.859   &  1417+385 & 15.4 & 1.831     \\
  0202+149 & 6.4 & 0.405   &  1458+718 & 3.907 & 0.904     \\
  0202+319 & 4.686 & 1.466   &  1502+106 & 10.2 & 1.839     \\
  0212+735 & 4.87 & 2.367   &  1504-166 & 3.413 & 0.876     \\
  0215+015 & 18.466 & 1.715   &   1510-089 & 16.11 & 0.360     \\
  0224+671 & 5.895 & 0.523   &  1546+027 & 8.5675 & 0.414      \\
  0234+285 & 12.12 & 1.207   & 1548+056 & 7.7 & 1.422     \\
  0333+321 & 12.2 & 1.259   &  1606+106 & 17.1 & 1.226     \\
  0336-019 & 13.07 & 0.852   & 1611+343 & 7.69 & 1.397     \\
  0420-014 & 5.6 & 0.914   &  1633+382 & 16.6625 & 1.814      \\
  0458-020 & 15.045 & 2.286   &  1637+574 & 9.07 & 0.751     \\
  0528+134 & 11.036 & 2.070   & 1638+398 & 8.266 & 1.666    \\
  0529+075 & 8.325 & 1.254   &  1641+399 & 12.4175 & 0.593     \\
  0529+483 & 17.54 & 1.162   &  1726+455 & 1.873 & 0.717     \\
  0552+398 & 0.363 & 2.363   &  1730-130 & 17.622 & 0.902     \\
  0605-085 & 16.186 & 0.872   & 1751+288 & 3.07 & 1.118     \\
  0730+504 & 12.75 & 0.720    &  1800+440 & 15.04 & 0.663     \\
  0736+017 & 9.332 & 0.191   &  1849+670 & 22.1 & 0.657    \\
  0738+313 & 6.986 & 0.631   &  1928+738 & 4.774 & 0.302    \\
  0748+126 & 14.365 & 0.889   &  1936-155 & 2.6 & 1.657     \\
  0804+499 & 1.83 & 1.436   & 1958-179 & 1.9 & 0.650     \\
  0827+243 & 17.7675 & 0.940   &  2037+511 & 3.3 & 1.686     \\
  0836+710 & 19.35 & 2.218   &  2121+053 & 10.845 & 1.941      \\
  0838+133 & 8.223 & 0.681   &  2136+141 & 3.7975 & 2.427      \\
  0906+015 & 19.645 & 1.024   &   2145+067 & 2.206 & 0.990     \\
  0923+392 & 2.44 & 0.695   &  2155-152 & 12.66 & 0.672     \\
  0945+408 & 13.256 & 1.249   &  2216-038 & 5.55 & 0.901      \\
  0955+476 & 2.48 & 1.882   &  2223-052 & 13.0425 & 1.404   \\
  1036+054 & 6.065 & 0.473   &  2227-088 & 8.1 & 1.560      \\
  1038+064 & 7.26 & 1.265    & 2243-123 & 3.88 & 0.632      \\
  1045-188 & 6.085 & 0.595   & 2251+158 & 7.27 & 0.859    \\
  1150+812 & 5.878 & 1.250   &  2331+073 & 3.445 & 0.401     \\
  1156+295 & 20.15 & 0.729    &  2345-167 & 11.21 & 0.576      \\ 
\hline
\multicolumn{6}{l}{Mean $\beta_{app}$ from \cite{Lister2009-2};
  redshifts from \cite{Lister2009}.}
\end{tabular}
\end{minipage}
\end{table*}

\begin{table*}
\setlength{\tabcolsep}{11pt}
 \centering
 \begin{minipage}{140mm}
 \centering
  \caption{BL Lac sample}
  \label{tab:BLLAC}
  \begin{tabular}{@{}ccccccc@{}}
  \hline
   Object Name & $\beta_{app} $ & Redshift  &  Object Name & $\beta_{app} $ & Redshift     \\
 \hline
 0003-066 & 2.145 & 0.347  &  1413+135 & 0.755 & 0.247     \\
   0716+714 & 10.07 &  0.310   &  1538+149 & 4.525 & 0.605      \\
   0754+100 & 14.4 &  0.266   & 1749+096 & 4.013 & 0.322      \\
  0808+019 & 13.000 & 1.148    &  1803+784 & 2.396 &  0.680     \\
  0814+425 & 1.060 & 0.245   &  1807+698 & 0.056 & 0.051     \\
   0823+033 & 14.3 & 0.506    &  1823+568 & 7.225 & 0.664     \\
  0829+046 & 6.347 & 0.174   &   2131-021 & 10.318 & 1.285      \\
   0851+202 & 8.613 & 0.306   &  2200+420 & 5.1335 & 0.068      \\
\hline
\multicolumn{6}{l}{Mean $\beta_{app}$ from \cite{Lister2009-2};
  redshifts from \cite{Lister2009}. }
\end{tabular}
\end{minipage}
\end{table*}

\section{Model Optimization}\label{parameters}

\subsection{Observables and Model Acceptability Criteria}\label{acceptability}

Sources in our adopted sample have available measurements for each of
the following quantities:
 redshift $z$; mean apparent jet speed $\langle \beta_{\rm app}
\rangle$; and average flux density $\langle S_\nu\rangle$.

In order for a model to be deemed
acceptable, we require that it adequately reproduce the observed
$z$ and $\langle \beta_{\rm app} \rangle$ distributions, when the
appropriate flux limit is applied. Formally, we require a
Kolmogorov-Smirnov test to return a probability higher than 5\% that
the simulated and observed values of $z$ (and, similarly, of $\langle
\beta_{\rm app} \rangle$) are drawn from the same
distribution. 

In contrast, as discussed in \S \ref{model}, we do not require the model to reproduce the observed
distribution of $\langle S_\nu \rangle$. We compare, however, the observed flux density distribution of the MOJAVE sample
\citep{Lister2009} and the flux density distribution of the optimal
model for the BL Lac objects and the FSRQs and discuss their agreement
in \S \ref{fllum}. 
A detailed analysis of the effect of variability on the flux density
distribution and the resulting fitted luminosity function will be the
subject of a future publication. 

The procedure for testing the acceptability of a specific model is the
following: 
\begin{itemize}

\item We use redshift bins of size 0.1 from $z=0$ to $z=1.4$ for
  the BL Lac sample and from $z=0$ to $z=2.5$ for the FSRQ
  sample. Since dN$\propto$dV for each redshift bin, we calculate the
  comoving volume element and the number of repetitions performed is
  proportional to dV. We also calculate the proper distance for each
  redshift and the look-back time in order to adjust the luminosity
  limits in each redshift bin.

\item For each repetition (simulated source), we randomly choose a
  value for $\cos(\theta)$, $\Gamma$, and $L_\nu$ according to
  the corresponding distributions of the specific model being tested.
 
\item From these values we calculate the velocity $\beta$, the apparent velocity $\beta_{app}$, the Doppler factor $D$ and the flux density $S_\nu$.

\item In order to simulate a flux-limited sample, we discard any
  source with $S_\nu$ lower than 1.5 Jy.

\item We construct the cumulative distribution function of the
  apparent velocity and redshift distributions of simulated sources,
  and we compare them to those obtained from the data. We use the
  Kolmogorov-Smirnov test (K-S test) in order to obtain the
  probability of the observed and the simulated data sets having been drawn
  from the same distribution.   
\end{itemize}

\subsection{A newly optimized model}\label{new opt model}

The parameters we optimize for each population are the slope $A$ of the luminosity
function, the evolution parameter $\tau$, and the slope $\alpha$ of
the Lorentz factor distribution. 
 We refer to the literature \citep{Lister1997} for an estimate of the
 power law indices for the luminosity and Lorentz factor distributions
 as a starting point, and proceed to explore the parameter space
 setting the lower limit for the luminosity to $10^{24}WHz^{-1}$ and
 the upper limit $10^{27}WHz^{-1}$ \citep{Arshakian2006}. A summary of
parameter values we adopt from the literature or directly from
the extrema of the datasets are shown in the upper part of Table
\ref{tab:Parameters}. 

We first perform a coarse preliminary scan of the parameter
space to derive initial values for the parameters to be optimized, starting from the aforementioned literature values and shifting towards higher or lower values according to the K-S test and visual inspection of the probability density and cumulative distribution functions.  
 During that investigation we have come to the conclusion that
the luminosity function of the BL Lacs does not evolve with redshift,
consistent with the recent findings of \citet{Ajello2014}: the pure
luminosity evolution model with a finite evolution parameter was
unable to adequately describe the redshift distribution of the BL Lac
objects. All the corresponding K-S test of the BL Lac redshift distribution with an evolution parameter close to the literature value gave probabilities $\leq 10^{-5}\%$ of consistency. The probability values increased while increasing the value of the evolution parameter, and reached acceptable levels, as described in \S \ref{acceptability}, when the evolution parameter was set to infinity.

Starting from the initial values obtained in our coarse scan, we
optimize the model parameters in the following way. We create a
3-dimensional cube in parameter space for the FSRQS ($A,\tau,\alpha$)
and a 2-dimensional plane for the BL Lacs ($A,\alpha$). Each side of
the cube is centered in the corresponding parameter value generating the smallest
inconsistency between observed and simulated cumulative distribution
functions, and its extent is equal to the coarse-scan step. 

\begin{table}
\setlength{\tabcolsep}{11pt}
\centering
  \caption{Model Parameters. Upper part: model values adopted from the
  literature or directly from extrema of observable distributions (see
text). Lower part: optimal parameter values; spread represents 
scanning step and statistical variations in simulated distributions. The second asymmetrical errors represent the range within which a parameter can produce "acceptable models"}
  \label{tab:Parameters}
\begin{tabular}{@{}cccc@{}}
 \hline
      & BL Lacs &  FSRQs  \\
  \hline
   $\Gamma_{min}$  &  1  & 1 \\
  $\Gamma_{max}$ &  16  & 26 \\
  $L_{min} ({\rm W\,Hz^{-1}})$ & $10^{24}$  & $10^{24}$ \\
  $L_{max} ({\rm W\,Hz^{-1}})$ & $10^{27}$  & $10^{27}$  \\
  s & 0 & 0.37 \\
  \hline
  $\alpha$ & 0.738$\pm 0.002^{+0.41}_{-1.46}$ & 0.57$\pm 0.001^{+0.12}_{-0.50}$ \\
  & & \\
  A & 2.251$\pm 0.02^{+0.68}_{-0.78}$ & 2.6$\pm 0.01^{+0.185}_{-0.245}$  \\
  & & \\
  $\tau(1/H_0)$ & - & 0.26$\pm 0.001^{+0.068}_{-0.003}$ \\
\hline
\end{tabular}
\end{table}

\begin{figure}
\resizebox{\hsize}{!}{\includegraphics[scale=1]{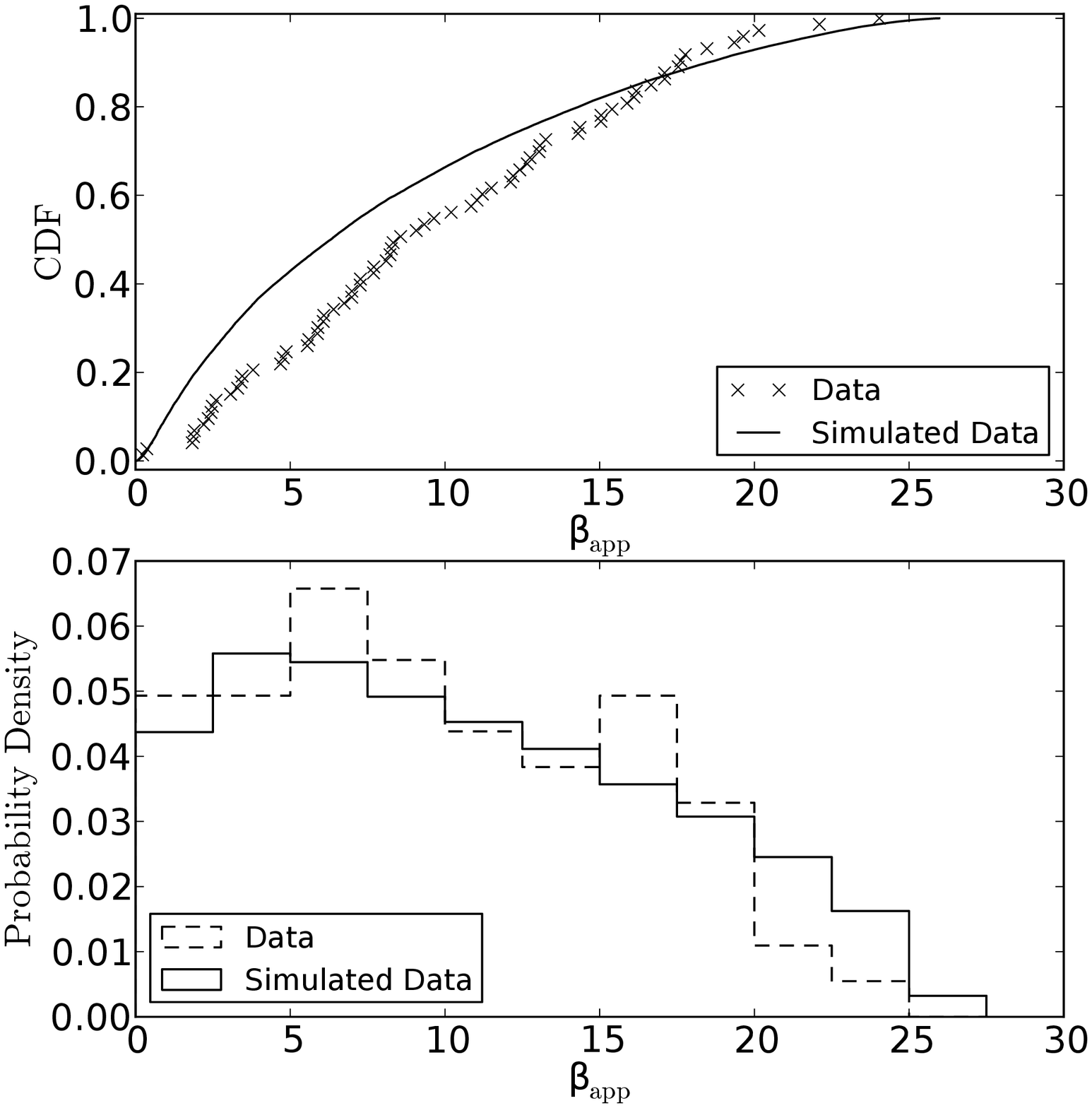} }
 \caption{Cumulative distribution function (upper panel) and
   probability density function (lower panel) of simulated (with our
   optimal model) and observed (MOJAVE sample, see text) $\beta_{app}$ for FSRQs.}
 \label{cpdf_bapp_qso}
 \end{figure}

\begin{figure}
\resizebox{\hsize}{!}{\includegraphics[scale=1]{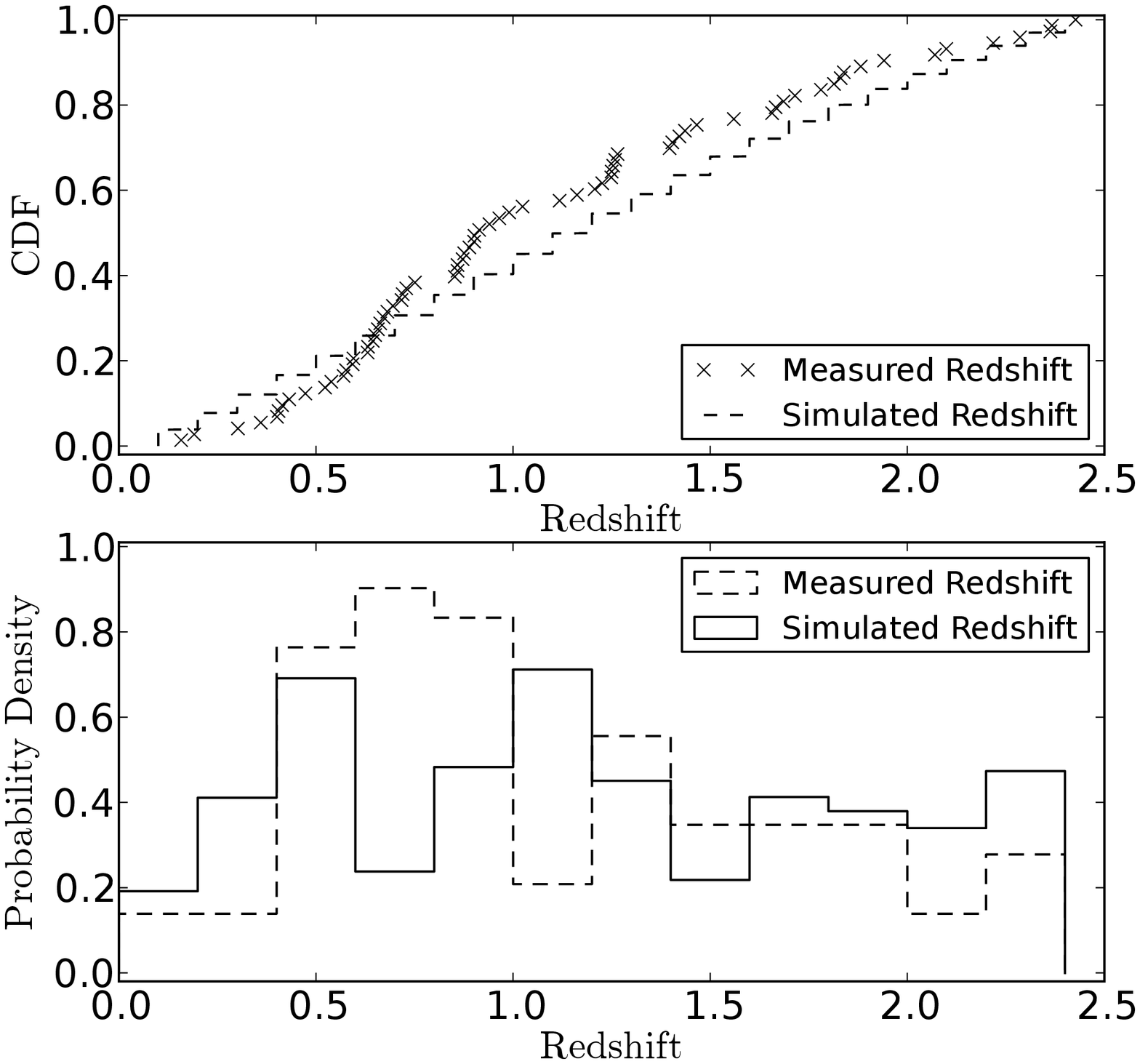} }
 \caption{Cumulative distribution function (upper panel) and
   probability density function (lower panel) of simulated (with our
   optimal model) and observed (MOJAVE sample, see text) redshifts $z$ for FSRQs}
 \label{cpdf_redshift_qso}
 \end{figure}

\begin{figure}
\resizebox{\hsize}{!}{\includegraphics[scale=1]{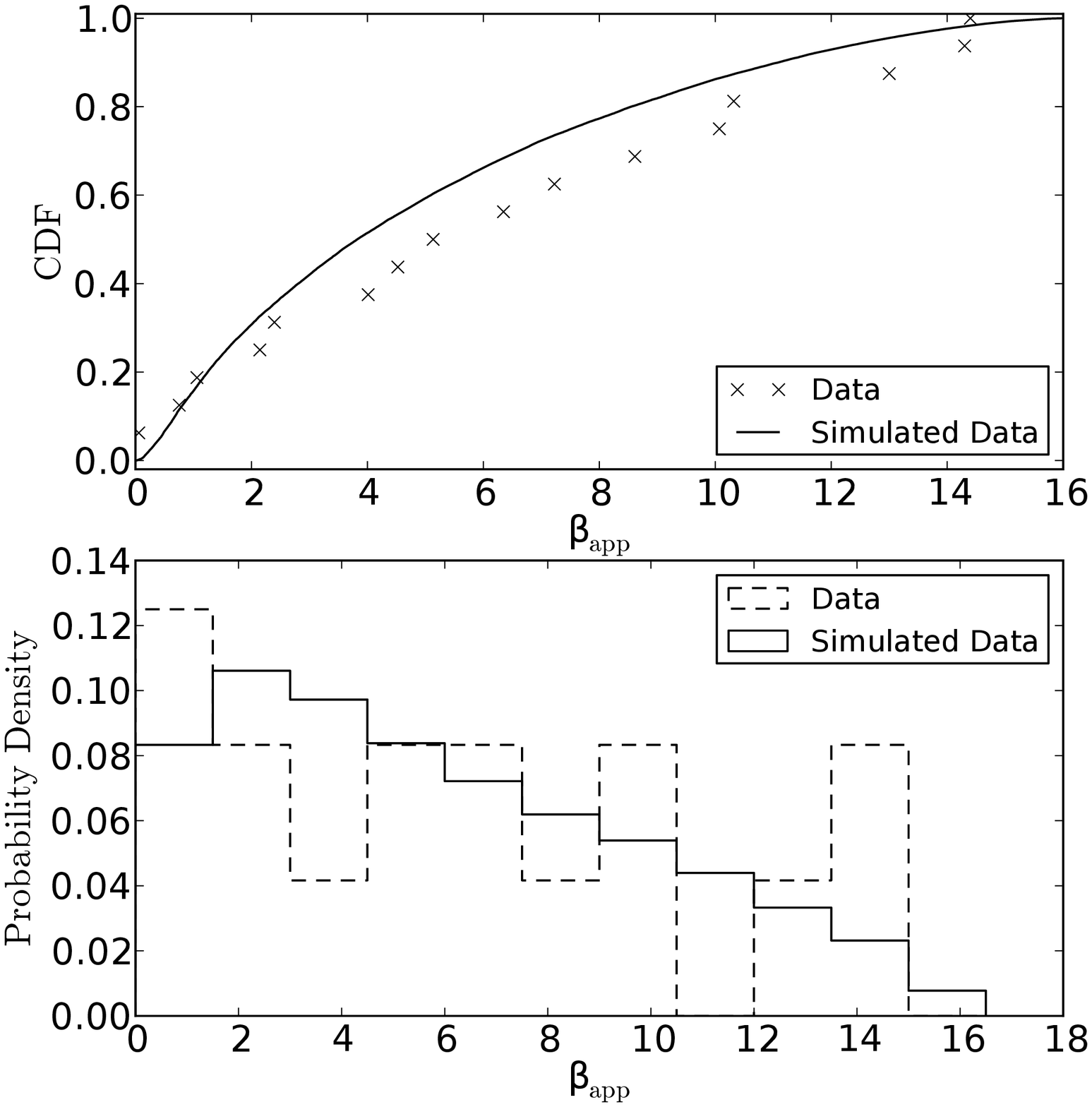} }
 \caption{Cumulative distribution function (upper panel) and
   probability density function (lower panel) of simulated (with our
   optimal model) and observed (MOJAVE sample, see text) $\beta_{app}$
   for BL Lacs}
 \label{cpdf_bapp_bllac}
 \end{figure}

\begin{figure}
\resizebox{\hsize}{!}{\includegraphics[scale=1]{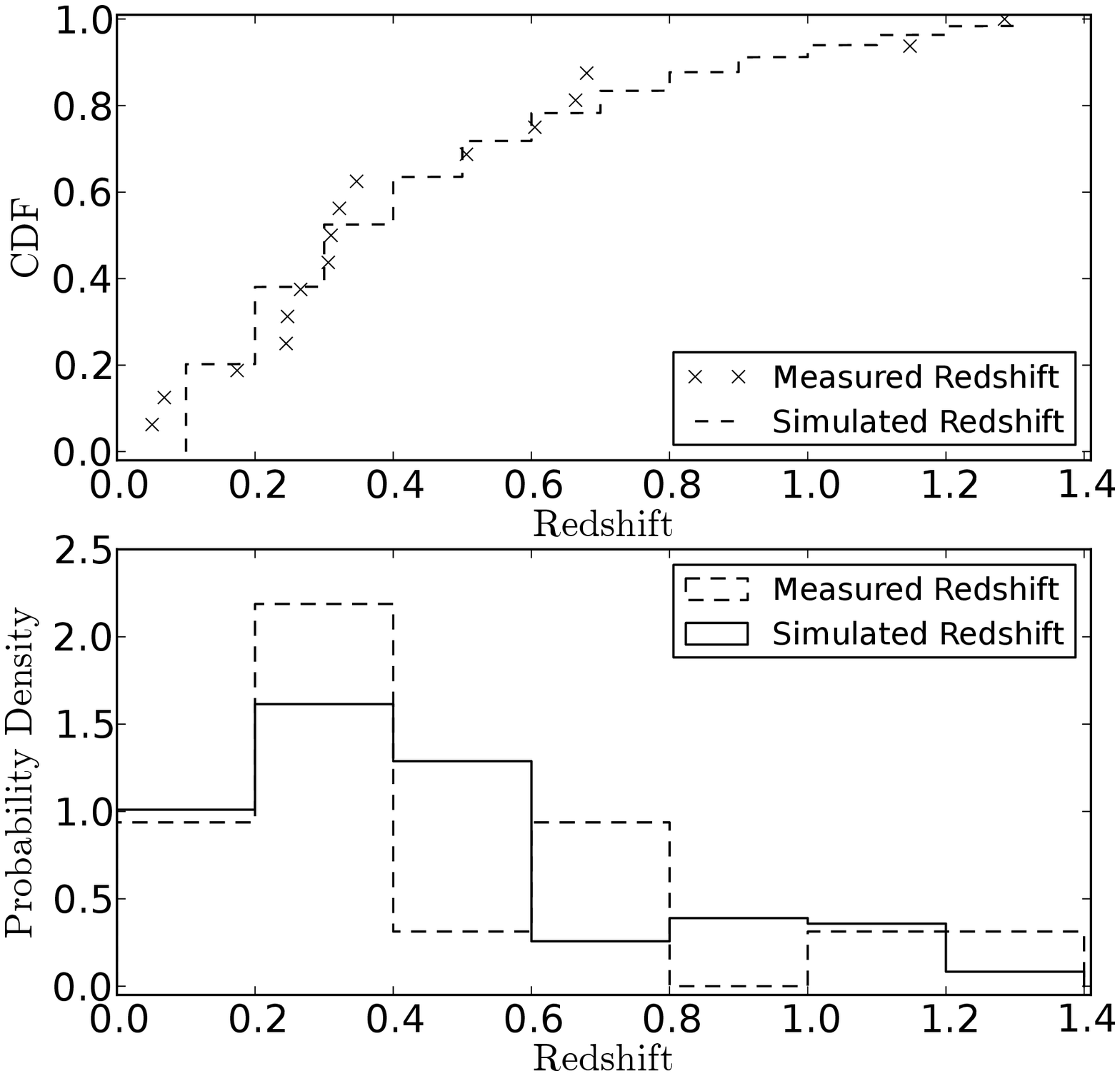} }
 \caption{Cumulative distribution function (upper panel) and
   probability density function (lower panel) of simulated (with our
   optimal model) and observed (MOJAVE sample, see text) redshifts $z$
   for BL Lacs.}
 \label{cpdf_redshift_bllac}
 \end{figure}

For each sample, we take every possible combination of parameters and
choose the one that minimizes the combined inconsistency between the observed
and simulated cumulative
distribution functions of the $\beta_{app}$ and redshift distributions
(we quantify that ``combined inconsistency'' by the product of the
corresponding K-S test statistics). 
We then repeat the process three additional times, adjusting the
boundary values to the step of the previous scan, and using smaller
steps centered around the previously obtained optimal parameters, in
order to converge to the best result. 
In order to investigate the effects of random sampling and account for
statistical deviations, we repeat several times the final scan. 
The optimal parameters values are shown in the lower part of
Table \ref{tab:Parameters}. Values and error bars correspond to the average
and the spread (standard deviation) of the various incarnations of the
final-scan optimization. They are
are therefore representative of the statistical spread in simulated
distributions (and indicative of the step in our final scan). In
contrast, the range of parameters that produce simulated distributions consistent with the
observable ones is addressed later in this section. 
For the FSRQ optimal parameters, the K-S test p-values are 49.3\% for
the apparent velocity and 8.4\% for the redshift distributions
(see also Fig. \ref{cpdf_bapp_qso},\ref{cpdf_redshift_qso}). For the BL Lac
optimal parameters, the corresponding K-S test p-values of 93.4\% for
the apparent velocity and 54.1\% for the redshift distributions (see
also Fig. \ref{cpdf_bapp_bllac},\ref{cpdf_redshift_bllac}).

It should be noted that the optimization procedure does not formally
correspond to proper model fitting that could be achieved, for
example, using a maximum-likelihood analysis, or a chi-square
minimization of the simulated probability density functions on the
observed ones. There are two reasons we have instead adopted the simpler procedure
described here and which is based on the K-S statistic. The first one
is simplicity. As discussed in \S \ref{model}, systematic
uncertainties in the observables and their interpretation are expected
to be significant, and affect the optimal model parameters much more
than the details of the optimization procedure. For this reason, the
investment in algorithmic complexity and computational time will
likely not yield a proportional improvement in model accuracy. The
second reason is that the K-S statistic provides an automatic way to
assess model consistency with the data: a nominal fitting procedure
will yield the model parameters {\em in best agreement} with the data, without
any a priori guarantees that this family of model is a {\em good} (or
even acceptable) description of
that dataset. Our procedure however allows us to automatically reject
poor fits up to a desired K-S p-value level and find the range of
parameters (if such a range exists) for which the hypothesis that the
chosen family of models produces the observed dataset cannot be rejected. 
 
In order to determine that range that produces ``acceptable models''
for each parameter,  we create a 2-D (for the BL Lacs), and a 3-D (for the FSRQs) parameter space centered around the best-fit values for each model. We use Monte Carlo sampling of the parameter space, to test which combination of parameters produce "acceptable" models, i.e. the K-S test yields a > 5\% probability of consistency between observed and simulated distributions.  We repeat this process enough times in order to ensure that the parameter space is adequately sampled. The range of parameter values produced in this way represents the spread of each parameter. If the extrema of these values coincide with the extrema of the initial  parameter space, we increase the range of that space and repeat the process. For the FSRQs this procedure yields $\alpha=0.57^{+0.12}_{-0.50}$, $A=2.6^{+0.185}_{-0.245}$ $\tau=0.26^{+0.068}_{-0.003}$. For the BL Lacs the procedure yields $\alpha=0.738^{+0.41}_{-1.46}$, $A=2.251^{+0.68}_{-0.78}$.

\section{Results}\label{results}

\subsection{BL Lacs} \label{BL}

\begin{figure}
\resizebox{\hsize}{!}{\includegraphics[scale=1]{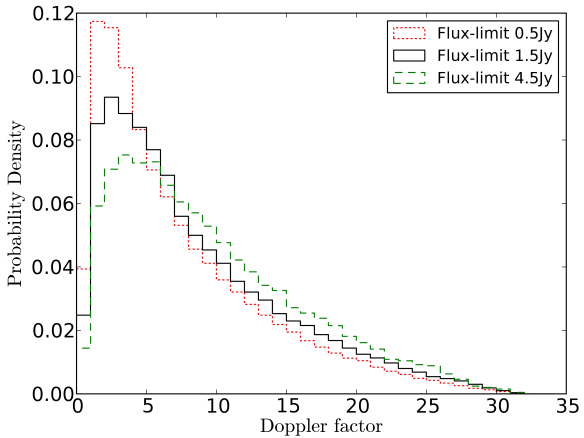} }
 \caption{BL Lac Doppler factor distribution. Different line types
   corresponds to different flux limits as follows. Solid
   line: 1.5 Jy; dashed line: 4.5 Jy; dotted line: 0.5 Jy.}
 \label{plt_pdfdoppler_bllac}
 \end{figure}
 
\begin{figure}
\resizebox{\hsize}{!}{\includegraphics[scale=1]{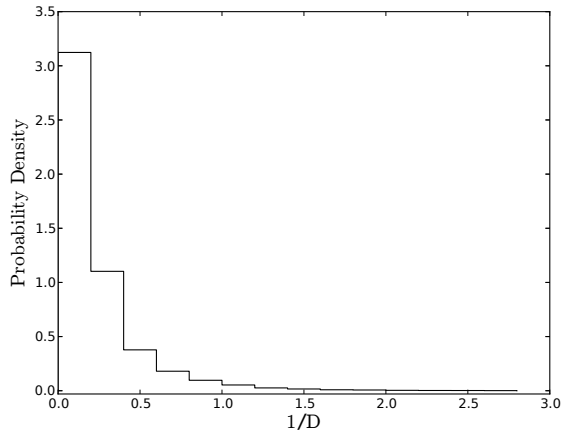}}
 \caption{Inverse Doppler factor distribution for a flux limited
   ($S_\nu \geq$1.5Jy) BL Lac sample.}
 \label{plt_pdfinvdoppler_bllac}
 \end{figure} 
 
\begin{figure}
\resizebox{\hsize}{!}{\includegraphics[scale=1]{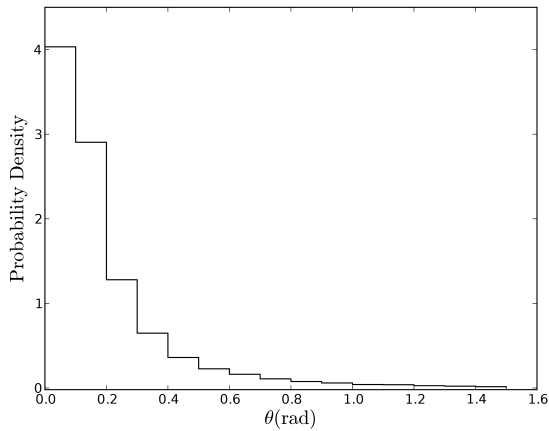}}
 \caption{Distribution of the viewing angles $\theta$(rad) for a flux limited
   ($S_\nu \geq$1.5Jy) BL Lac sample. The 
   majority of simulated sources have small viewing angles ($68.8\%$
   have $\theta\leq 0.2 {\rm \, rad}=11^.5\circ$), in accordance with
our understanding of blazars.}
 \label{plt_pdftheta_bllac}
 \end{figure}

\begin{figure}
\resizebox{\hsize}{!}{\includegraphics[scale=1]{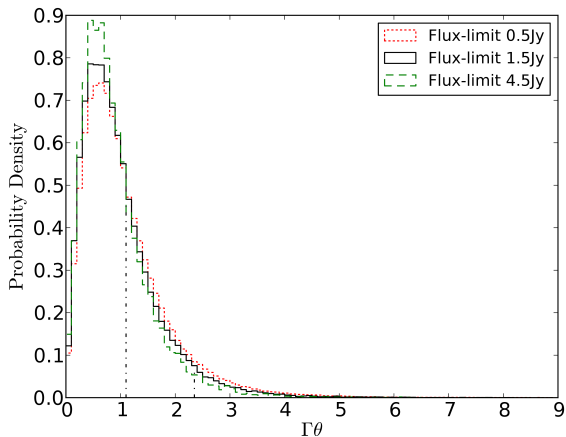}}
 \caption{Distribution of $\Gamma\theta$ for flux-limited BL Lac
   samples. Different lines correspond to different flux limits as
   follows. Solid line: 1.5 Jy; dashed line:4.5 Jy; dotted line: 0.5
   Jy. The vertical dashed line at 1.11 represents $1\sigma$, at 2.37
   $2\sigma$ and the dashed line (not visible at $\Gamma\theta=4.8$)
   $3\sigma$ for the 1.5 Jy distribution.} 
 \label{plt_pdfthlor_bllac}
 \end{figure}
 
 \begin{figure}
\resizebox{\hsize}{!}{\includegraphics[scale=1]{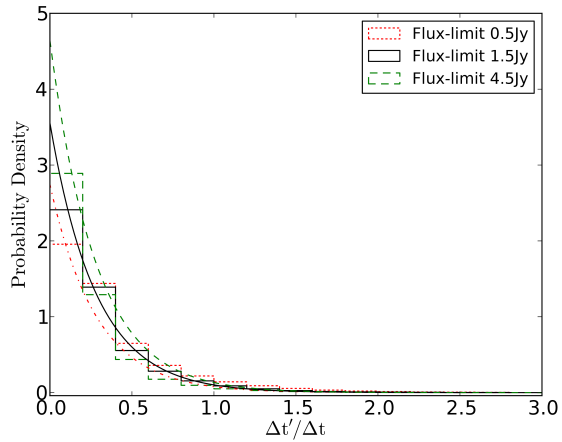}}
 \caption{Time scale modulation factor distribution for BL Lac
   objects. Different line types correspond to different flux limits
   as follows. Solid line:1.5 Jy; dashed line:4.5 Jy; dotted line: 0.5
   Jy. The overploted smooth lines represent exponential distributions with
   the same mean as each histogram.}
 \label{plt_pdftscale_bllac}
 \end{figure}

Using the optimal values for the two parameters of our BL Lac
population model, we explore, in this section, 
the distributions of the derived quantities discussed in \S \ref{model}. The Doppler factor
distribution and inverse Doppler, which will be important in the
application to the timescale analysis, are shown in
Fig. \ref{plt_pdfdoppler_bllac} \& \ref{plt_pdfinvdoppler_bllac}. The
distribution of the viewing angles for sources that pass the flux
limit is shown in
Fig. \ref{plt_pdftheta_bllac}, and the distribution of the product
($\Gamma \theta$) in
Fig. \ref{plt_pdfthlor_bllac}. 

We can clearly see in Fig.~(\ref{plt_pdftheta_bllac}) that the sources
that manage to pass the flux limit are strongly biased towards very
small viewing angles, consistent with our
understanding of blazars having jets closely aligned with our line of sight.
Although most sources in the inverse Doppler factor distribution have
$1/D<1$ (Fig.~\ref{plt_pdfdoppler_bllac}), there is a non-zero power
at $1/D>1$. All sources with $1/D>1$ are not boosted. 

The optimal parameters presented in Table
\ref{tab:Parameters} are applicable to the entire BL Lac
distribution. In contrast, 
the resulting derivative-quantity distributions presented in this
section are specific to a sample (characterized by a certain flux
limit). To demonstrate this explicitly, we plot, 
in Figs. \ref{plt_pdfdoppler_bllac} and \ref{plt_pdfthlor_bllac}, the
resulting distributions ($D$ and $\Gamma \theta$) if we implement a different
flux limit. We choose to use limits a factor of 3 higher (4.5 Jy) and
lower (0.5 Jy) than the limit of the observed dataset (1.5 Jy). In
both cases we see that the distributions have similar shapes, and
the location of their peaks are insensitive to the flux limit;
however, the power in the tails compared to the peak changes as the value
of the flux limit changes. As expected, the Doppler factor distribution has more
power in the tails (a larger fraction of highly boosted sources) as
the flux limit increases. The $\Gamma\theta$ distribution shows an
increase in the number of sources with lower values (more highly
beamed sources) as the flux limit increases. We therefore emphasize that
although our optimized $\Gamma$ and $L_\nu$ distributions describe
blazars as a population, the correct flux limit 
must be taken into account when using them to obtain distributions of derivative
quantities to compare with a specific flux-limited sample. 

It is common in studies of beamed sources, when there is not enough
information to estimate both $\Gamma$ and $\theta$ for a source, to
assume an average value of $\Gamma \theta$ (typically,
$\Gamma\theta=1$, e.g. \citealp{Vermeulen1995,Cohen2007}). The
preference for $\Gamma \theta = 1$ stems from the fact that, if the Lorentz factor associated with the
observed  movement of the knots in a jet is also the Lorentz factor of
the bulk flow, then the maximum apparent speed for a given $\Gamma$ is
achieved for $\sin\theta=1/\Gamma$. Thus for small
angles $\theta\sim 1/\Gamma$. 
Figure \ref{plt_pdfthlor_bllac} allows us to evaluate the validity of
that assumption. The most likely value of $\Gamma \theta$  is at 0.6
(see also \citealp{Jorstad2005}), and this result is quite robust
with respect to the sample flux limit. The mean for $S_\nu \geq$ 1.5 Jy
is 0.95, close to the frequently assumed $\Gamma \theta$
value. However, we point out that the spread of the $\Gamma \theta$
distribution is large: $68 \%$ of all values are included between
$0\leq \Gamma \theta \leq 1.1$. For this reason, any assumption
regarding the value of $\Gamma \theta$, whether that value is the mean
or the mode of the simulated distributions, should be treated with
caution. 

The emission from a region with bulk relativistic Lorentz factor $\Gamma$ is generally
beamed within a cone of opening angle $1/\Gamma$. For this reason, for beamed sources we expect $\Gamma\theta<1$. Due to the spread of the distribution, we conclude that most, but not all, of the sources in a flux-limited sample of relativistic jets are beamed. 

Combining the Doppler factor and redshift distributions
(Eq.\ref{eqtscale}) we derive the timescale modulation factor
distribution (Fig. \ref{plt_pdftscale_bllac}). The distribution can be
well described by an exponential distribution with mean equal to the
data mean. In the case of a flux limit of 1.5 Jy, the mean
timescale modulation factor is equal to 0.281 . However, the
distribution is not a strict exponential, because the maximum Doppler
factor is finite and hence the time modulation factor is never exactly
equal to zero. Its smallest value in the $S_\nu \geq 1.5 Jy$ sample is
$\sim 3\times 10^{-2}$. In Fig. \ref{plt_pdftscale_bllac}  we also plot the resulting
distributions if we implement a different flux limit, as in 
Figs.~\ref{plt_pdfdoppler_bllac} and \ref{plt_pdfthlor_bllac}. Higher
flux limits result to a larger fraction of sources with very
compressed timescales.

\subsection{Flat spectrum Radio Quasars}\label{QSO}

\begin{figure}
\resizebox{\hsize}{!}{\includegraphics[scale=1]{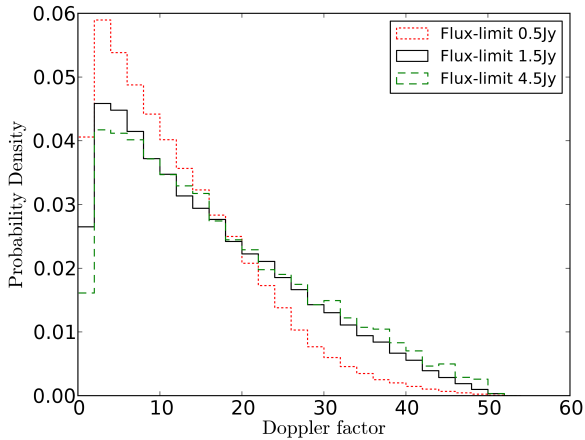}}
 \caption{Doppler factor distribution for FSRQs. Different
   line types correspond to different flux limits, as follows. Solid
   line:1.5 Jy; dashed line: 4.5 Jy; dotted line: 0.5 Jy.}
 \label{plt_pdfdoppler_qso}
 \end{figure}
 
\begin{figure}
\resizebox{\hsize}{!}{\includegraphics[scale=1]{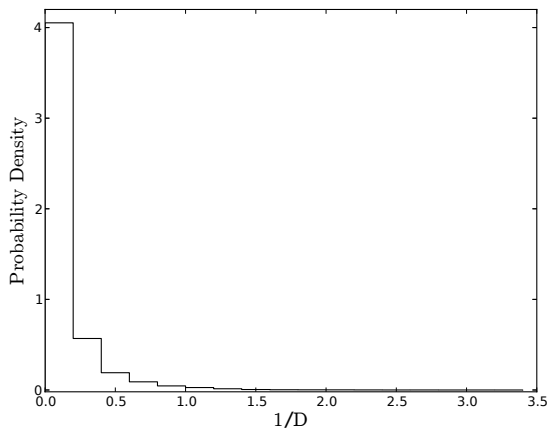}}
 \caption{Inverse Doppler factor distribution for the FSRQs sample.}
 \label{plt_pdfinvdoppler_qso}
 \end{figure}

\begin{figure}
\resizebox{\hsize}{!}{\includegraphics[scale=1]{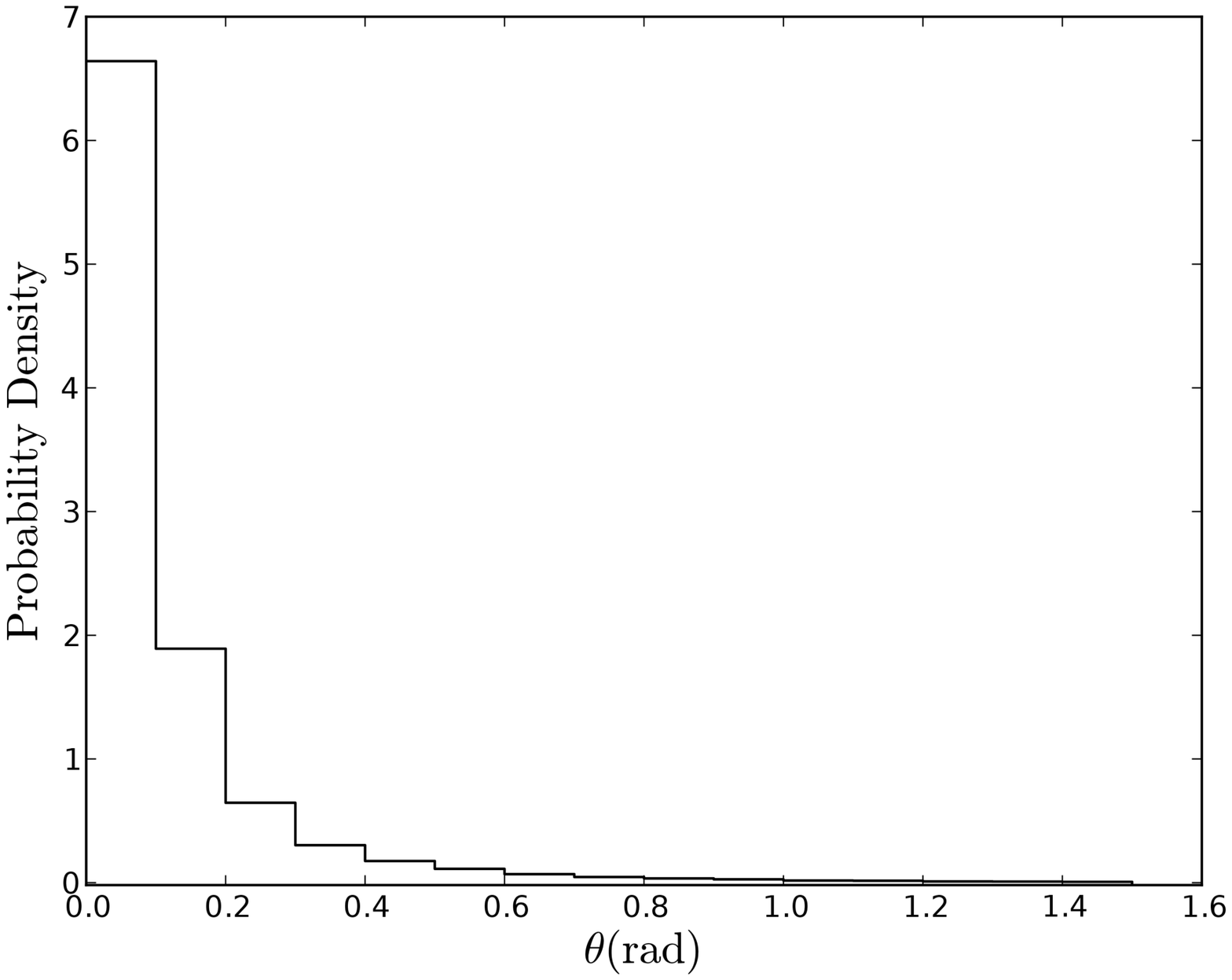}}
 \caption{Distribution of the viewing angles $\theta$(rad). The 
   majority of simulated sources have small viewing angles ($85\%$
   have $\theta\leq 0.2 {\rm \, rad}=11.5^\circ$), in accordance with
our understanding of blazars. 
}
 \label{plt_pdftheta_qso}
 \end{figure}

\begin{figure}
\resizebox{\hsize}{!}{\includegraphics[scale=1]{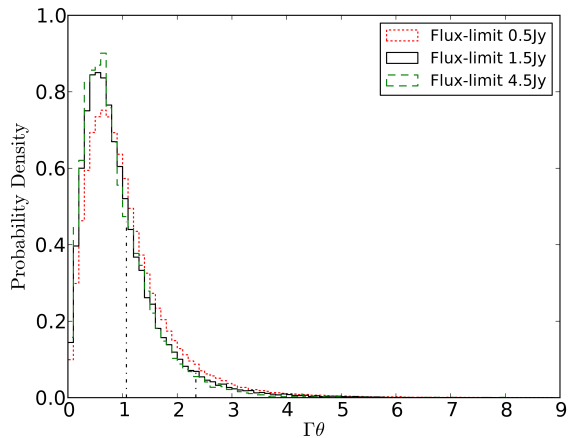}}
 \caption{Distribution of $\Gamma\theta$ for the FSRQ sample. The
   solid line represents the distribution with the flux limit we set
   for our model (1.5 Jy). The dashed line is the distribution fora
   flux limit of 4.5 Jy and the dotted line for a limit of 0.5 Jy. The vertical dashed line at 1.07 represents $1\sigma$, at 2.34 $2\sigma$ and the dashed line (not visible at $\Gamma\theta=5.0$) $3\sigma$ for the 1.5 Jy distribution. }
 \label{plt_pdfthlor_qso}
 \end{figure}

\begin{figure}
\resizebox{\hsize}{!}{\includegraphics[scale=1]{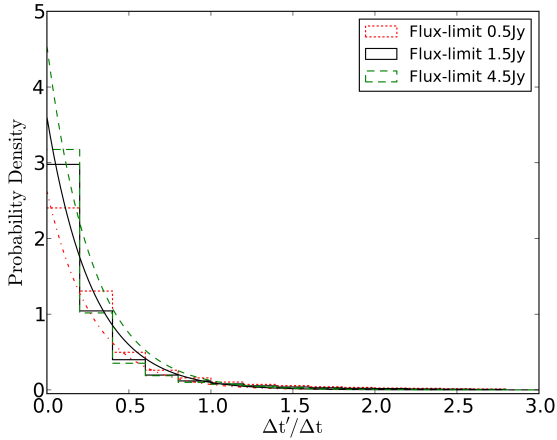}}
 \caption{Time scale modulation factor distribution of the simulated
   data for the FSRQ sample. The solid line represents the
   distribution with the flux limit we set for our model (1.5 Jy). The
   dashed line is the distribution for a flux limit of 4.5 Jy and the dotted line for a limit 0.5 Jy. The overploted lines represent exponential distributions with the same mean.}
 \label{plt_pdftscale_qso}
 \end{figure}

In this section, we discuss the same derived distributions for FSRQs
as the ones we derived for BL Lacs in \S \ref{BL}. The Doppler and inverse Doppler factor distribution and the distribution of the viewing angles are shown in Figs. \ref{plt_pdfdoppler_qso},\ref{plt_pdfinvdoppler_qso} and \ref{plt_pdftheta_qso} respectively. The Doppler factor distributions of the BL Lac and FSRQ samples show remarkable similarities. Even though FSRQs have higher Doppler factors, both distributions peak at ${\sim}3$ and have similar shapes.

The distribution of the ($\Gamma\theta$) is shown in
Fig. \ref{plt_pdfthlor_qso}. As in the case of BL Lacs, we find the
$\Gamma\theta$ to be peaked around 0.5, in agreement with the early analytic
predictions of \citet{Vermeulen1994} for quasars. The consistency
between the $\Gamma \theta$
distributions in FSRQs and BL Lacs is remarkable; differences in the
amount of relativistic beaming are not likely to be the culprit of any
observed differences between these two classes of sources. 

The same conclusion can be reached by comparing the time scale
modulation factor distribution for the FSRQs, shown in
Fig. \ref{plt_pdftscale_qso}, with that of BL Lacs. The time scale
modulation factors of a 1.5Jy flux-limited FSRQ sample also follow an
exponential distribution with mean=0.277. 
Similar to the BL Lacs, the smallest value is $\sim 2\times 10^{-2}$.

As for BL Lacs, we have also produced distributions for the Doppler
factor,$\Gamma\theta$ and the time scale
modulation factor for samples with
different flux limits. The effect of changing the flux limit is similar for the FSRQs as in the
case of BL Lacs. In all three cases the distributions have similar
peaks, and the Doppler factor distribution becomes shallower with
higher flux limit, whereas more sources have smaller values of
$\Gamma\theta$ and $\Delta t'/\Delta t$ for a higher flux limit.

\subsection{Flux density and Luminosity Distributions}\label{fllum}

\begin{figure}
\resizebox{\hsize}{!}{\includegraphics[scale=1]{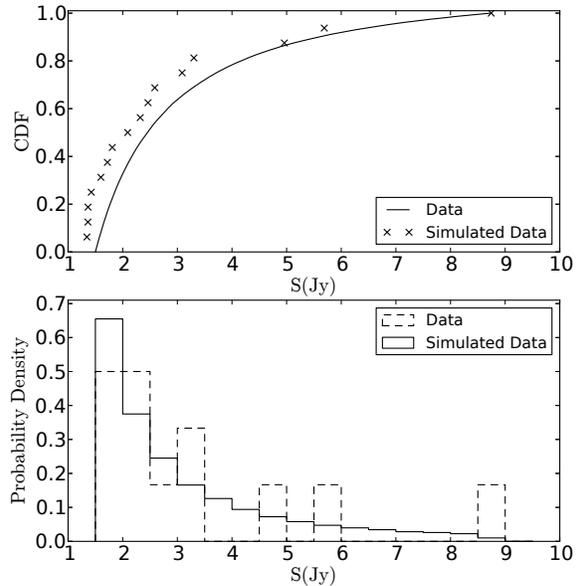}}
 \caption{Flux density distribution of the MOJAVE sample (BL Lacs) \citep{Lister2009}(dashed) and the simulated flux density (solid).}
 \label{plt_flux_bllac}
 \end{figure}

\begin{figure}
\resizebox{\hsize}{!}{\includegraphics[scale=1]{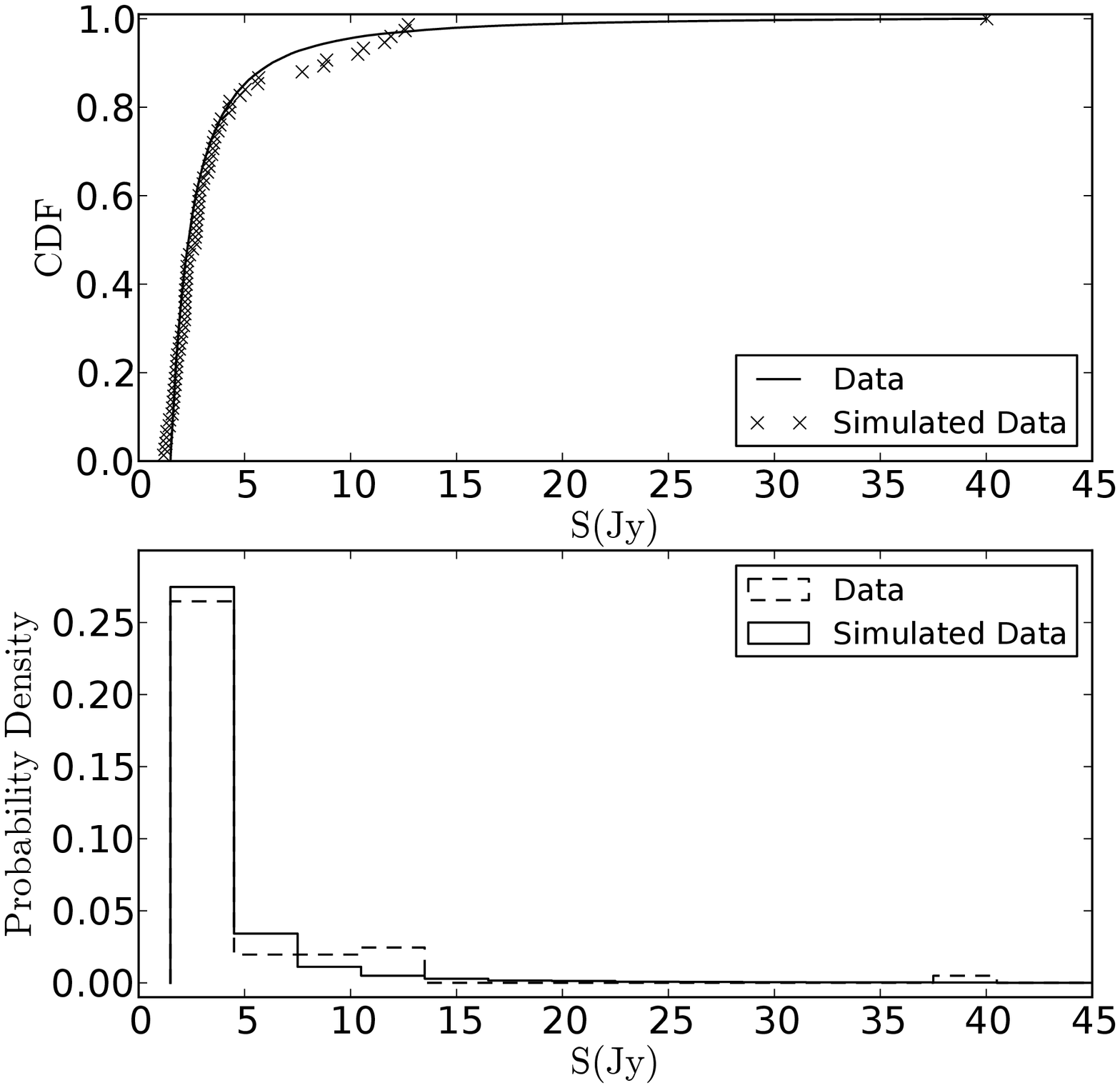}}
 \caption{Flux density distribution of the MOJAVE sample (FSRQs) \citep{Lister2009}(dashed) and the simulated flux density (solid).}
 \label{plt_flux_qso}
 \end{figure}  

\begin{figure}
\resizebox{\hsize}{!}{\includegraphics[scale=1]{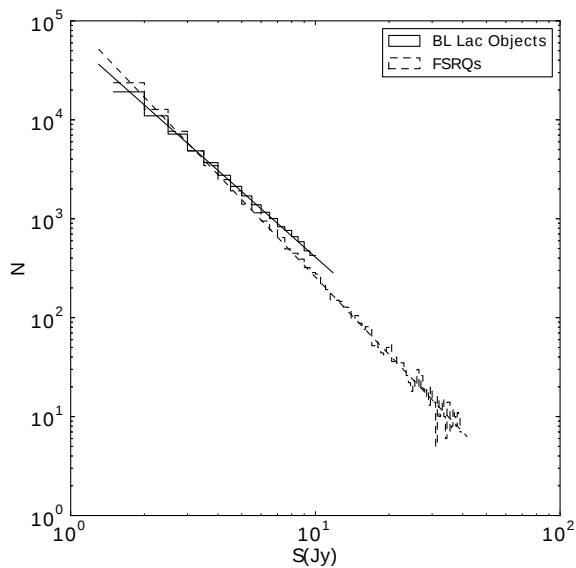} }
 \caption{$\log N$--$\log S_\nu$ plot for the FSRQ (dashed) and BL Lac
   (solid) 1.5 Jy - limited samples. The corresponding lines represent power law distributions with slope -2.2 for the BL Lac objects and -2.6 for the FSRQs.}
 \label{log_log_flux}
 \end{figure} 

\begin{figure}
\resizebox{\hsize}{!}{\includegraphics[scale=1]{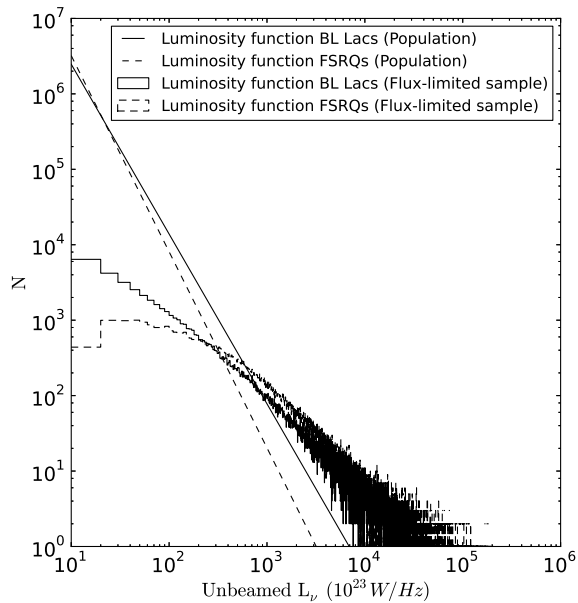} }
 \caption{Unbeamed $\log N$--$\log L_\nu$ for the FSRQ (dashed)
   and BL Lac (solid) 1.5 Jy - limited samples. The solid and dashed lines represent the input power law distribution at z=0 for the BL Lac and the FSRQ samples respectively.}
 \label{log_log_unbeamed_lum}
 \end{figure} 
 
\begin{figure}
\resizebox{\hsize}{!}{\includegraphics[scale=1]{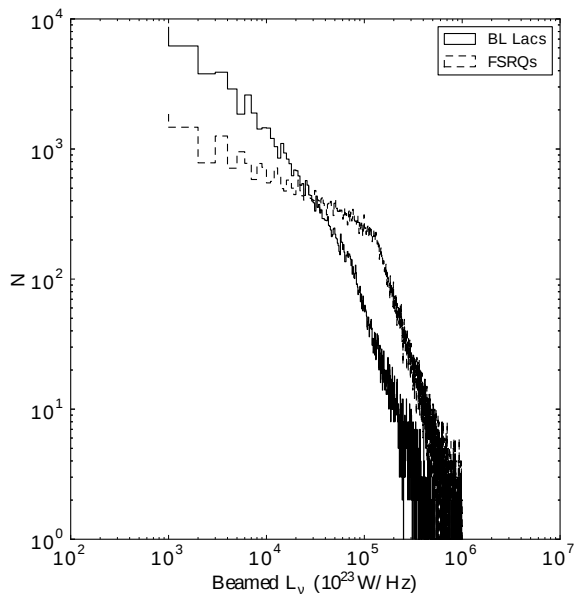} }
 \caption{Beamed $\log N$--$\log L_\nu$ for the FSRQ (dashed)
   and BL Lac (solid) 1.5 Jy - limited samples.}
 \label{log_log_beamed_lum}
 \end{figure}  
 
It was argued in \S \ref{acceptability} that due to the variable
nature of blazars, flux density is not a good observable for model
fitting. For this reason, we have not required that our acceptable
models produce a flux density distribution consistent with the data. 
It is nevertheless interesting to compare,
the flux density distributions in our simulated and observed 1.5Jy
flux-limited samples. 
Figures \ref{plt_flux_bllac} \& \ref{plt_flux_qso} show the
distribution of the derived flux density versus the highest flux
density from observations \citep{Lister2009} for the BL Lac and FSRQ
samples respectively. Despite our initial concerns regarding
variability, simulated and observed distributions do not
appear to be discrepant in either case. The Kolmogorov-Smirnov test
gives a probability of $\sim 23\%$
for the BL Lac objects and  $\sim 21\%$ for the FSRQs that the observed and simulated data are drawn
from the same distribution. 
A possible interpretation of this result is that the probability of
observing a source in a flaring state in a single-epoch survey  is low. \citet{Lister2001} has argued that
even though the flux densities of beamed sources can reach high levels
because of Doppler boosting, 
compression of flare timescales implies that these sources are most
likely to be observed relatively close to their quiescent
levels. 

Figure \ref{log_log_flux} shows the flux density distribution of
the two samples in the $\log N$ -$\log S_\nu$ format. The flux density distribution of
the sources above the 1.5Jy flux limit follows a power law distribution with a slope of -2.2 for the BL Lacs and -2.6 for the FSRQs.  
For comparison, a uniformly distributed, single-luminosity
extragalactic population in a flat cosmology yields a slope of -2.5.

The $\log N$--$\log L_\nu$ plots for unbeamed and beamed luminosities of
the sources that pass the 1.5 Jy flux limit are shown in Figs.
\ref{log_log_unbeamed_lum} and \ref{log_log_beamed_lum}
respectively. For the unbeamed case we have also overplotted the input
intrinsic luminosity functions at z=0. As expected, the intrinsic
distribution of the luminosity is stepper than the distribution of the
luminosity of the sources above the flux limit for both samples
(e.g. \citealp{CL08b}). 

In the case of the beamed luminosity functions we see a clear break in
the beamed $\log N$ -$\log L_\nu$ of the FSRQ sample at $\sim
2{\times}10^{28} {\rm W\,Hz^{-1}}$ and a less pronounced break in the
beamed $\log N$ -$\log L_\nu$ of the BL Lac sample.

The deviations from a single power law in the distribution of
luminosities {\em within the sample} are not a reflection of the
intrinsic luminosity function shape (which, in our case, is a single
power-law by construction). Rather, they are artifacts of beaming (see
also \citealp{Urry1984,Urry1991}) and mixing
objects from different redshifts; for example, the sources above and below the
break in the case of FSRQs are strongly dominated by objects at
redshifts higher and lower than $\sim 1$, respectively. 
However, past studies have shown that reconstruction of the luminosity
function starting from observations in a flux-limited sample can
erroneously map these breaks to the intrinsic luminosity function
shape \citep{Cara2008,CL08b}.  Our results do not show any need for a
blazar luminosity function shape more complex than a single power
law.

\section{Summary and Conclusions}\label{conclusions}

Using a Monte Carlo approach, we produced a new statistical model for
the blazar population, parameterized by the power law indexes of the
luminosity and Lorentz factor distributions, and the evolution
parameter. We derive distinct distribution parameters for FSRQs and BL
Lacs. Using this model we can
produce distributions for the Doppler factor, viewing angle, product $\Gamma \theta$,  and time scale modulation
factor for any flux-limited sample of beamed sources.

We have set out with the aim to answer the following three questions regarding the
blazar population. 

\subsection{What is the relativistically induced spread in observed event timescales?}

\begin{figure}
\resizebox{\hsize}{!}{\includegraphics[scale=1]{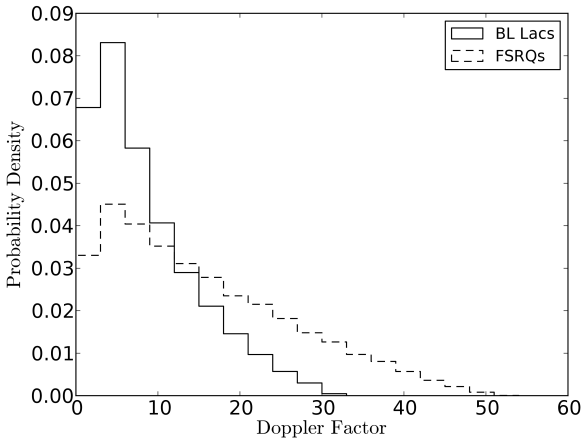}}
 \caption{Distributions of Doppler factors for the BL Lac (solid) and the FSRQ (dashed)
  1.5 Jy - limited samples} 
 \label{plt_doppler_joint}
 \end{figure}

Because of relativistic and cosmological expansion effects, even events with identical timescales
across sources would appear to have a distribution of observed
timescales, as long as the members of the sample we are considering
have a variety of Doppler factors (Fig. \ref{plt_doppler_joint}) and/or redshifts (Fig. \ref{cpdf_redshift_qso},\ref{cpdf_redshift_bllac}). 
We can quantify
this spread by considering the simple case where indeed we have a
class of events with identical rest-frame timescales $\tau_r=T$ across all sources, and
calculating the resulting distribution of observed timescales. 

Defining the timescale modulation factor $m=\Delta t'/\Delta t$,
$\tau_r$ is related to the observed timescale $\tau_o$ through
$\tau_o=m\tau_r$. The distribution of $\tau_r$ is a delta function, 
\begin{equation}
p(\tau_r) = \delta(\tau_r-T)\,.
\end{equation}
We have shown that the distribution of $m$ is an exponential, 
\begin{equation}
p(m) = C\frac{1}{m_0}\exp\left[-\frac{m}{m_0}\right], m_{\rm min} \leq
m \leq m_{\rm max}
\end{equation}
where $C$ is a normalization constant to account for the truncated
range. 
If $m$ and $\tau_r$ are independent, their joint probability function
is 
\begin{equation}
p(m,\tau_r)= C\frac{1}{m_0}\exp\left[-\frac{m}{m_0}\right]\delta(\tau_r-T)
\end{equation}
within the same limits as above. Transforming to new variables $m,
\tau_o$ and integrating over all values of $m$ we obtain
\begin{equation}
p(\tau_o) \propto \exp\left[-\frac{\tau_o}{T m_0}\right]\,.
\end{equation}

Inverting this problem, we conclude that the observation of an
exponential distribution of timescales associated with a class of
events in a flux-limited sample of BL Lacs or of FSRQs is consistent
with all timescales being identical in the rest frame of the jet. The
jet rest-frame timescale of such a class of events is equal to the
observed mean timescale divided by the average timescale modulation
factor of the relevant sample. 

\subsection{ Beaming of sources in flux-limited samples.} 

\begin{figure}
\resizebox{\hsize}{!}{\includegraphics[scale=1]{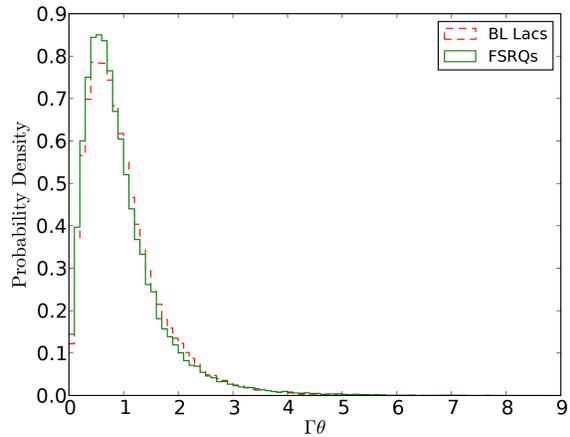}}
 \caption{Distributions of $\Gamma\theta$ for the BL Lac (dashed) and the FSRQ (solid) 1.5 Jy - limited samples} 
 \label{plt_thlor_joint}
 \end{figure}

In any flux-limited sample there is a well-defined peak
in the probability distribution of $\Gamma \theta$, and the location
of this peak is rather robust with respect to the value of the flux
limit (0.6 for BL Lacs, 0.5 for FSRQs Fig. \ref{plt_thlor_joint}). The mean of this distribution
is closer to 1 (0.99 for BL Lacs, 0.95 for FSRQs). As is indicated by
the significant difference between mode and mean, the distribution is
not only skewed but also its spread is large. Indeed, for a 1.5 Jy -
limited sample, 68\% of sources are contained in the interval 
$0 \leq \Gamma \theta \leq 1.11$, and 95\% of sources between 0 and 
2.37. The consequence is that, lacking enough information to compute
both $\Gamma$ and $\theta$ for a source, it is precarious to make a
statistical assumption for the value of $\Gamma \theta$ in order to
close the system and solve the problem. If any such assumption is
made, the $1\sigma$ uncertainty
places the value of $\Gamma \theta$ between 0 and 1.1, i.e. maximally
beamed and marginally beamed.  

For this reason, although the amount of beaming for a large sample of
sources can be usefully constrained through models of the type
discussed here, we strongly recommend against using such statistical
arguments on single sources. 

\subsection{ BL Lacs and FSRQs do NOT have different beaming properties.}

\begin{figure}
\resizebox{\hsize}{!}{\includegraphics[scale=1]{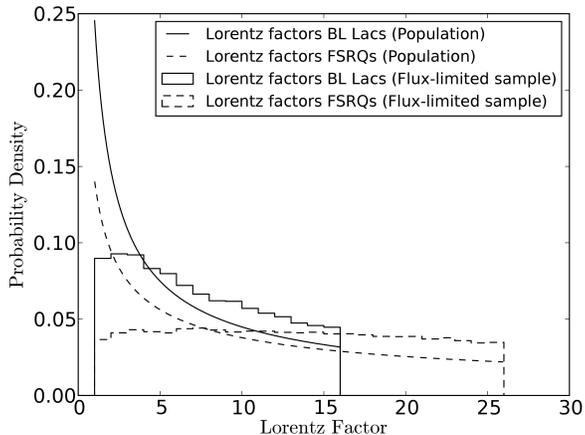}}
 \caption{Distributions of Lorentz factors for the FSRQ (dashed)
   and BL Lac (solid) 1.5 Jy - limited samples. The solid and dashed lines represent the input power law distribution  for the BL Lac and the FSRQ samples respectively.} 
 \label{plt_lorentz_joint}
 \end{figure}

\begin{figure}
\resizebox{\hsize}{!}{\includegraphics[scale=1]{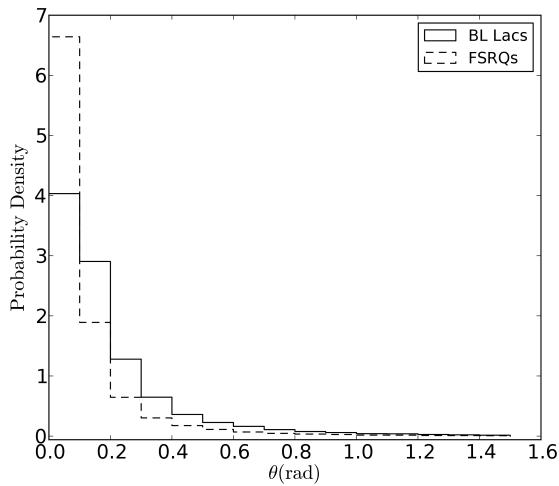}}
 \caption{Distributions of  viewing angles for the BL Lac (solid) and the FSRQ (dashed)
  1.5 Jy - limited samples. } 
 \label{plt_theta_joint}
 \end{figure}

 Although BL Lacs have a
steeper optimal slope in their $\Gamma$ distribution, lower values
of $\Gamma$ (Fig. \ref{plt_lorentz_joint}), and larger, on average, viewing angles (Fig. \ref{plt_theta_joint}), than FSRQs, the
distributions of the quantities characterizing beaming and timescale compression
($\Gamma \theta$ and $\Delta t'/\Delta t$, respectively), are very similar. 
The monoparametric exponential distributions that can describe the latter have
consistent mean: for the 1.5 Jy - limited sample and our optimal
model, these are 0.281 for the BL Lacs and 0.277 for the FSRQs. 

The consequence of this result is that any difference observed between
BL Lac and FSRQ flux-limited samples (with the same flux limit) in the time domain are not due to differences in
relativistic timescale compression between the two classes. Rather,
they must be reflecting an intrinsic difference between these two
classes of sources. 

There is some uncertainty in this result stemming from uncertainty in
our distribution parameters. In order to quantify this effect, we
calculate the mean of $\Delta t'/
\Delta t$ in a 1.5 Jy - limited sample for different values of the input distribution
parameters. As in \S \ref{new opt model}, we have kept all but one parameters
at their optimal values, and varied the remaining one to the limit
(maximum and minimum) where it still produces acceptable $z$ and $\beta_{\rm app}$
distributions (K-S test better than 5\%). The values of the limiting
parameter values are given in \S \ref{new
  opt model}, and the corresponding values of the $\Delta t'/
\Delta t$  means are given in
Table \ref{tab:time-scale modulation factor}. 
The largest possible deviation from our optimal parameter values is
produced for shallower luminosity function slopes in BL Lacs.

\begin{table}
 \setlength{\tabcolsep}{11pt}
\centering
  \caption{Model Parameters}
  \label{tab:time-scale modulation factor}
\begin{tabular}{@{}ccc@{}}
 \hline
   Parameters &$<\Delta{t'}/\Delta{t}>_{\rm BL Lacs}$  &
   $<\Delta{t'}/\Delta{t}>_{\rm FSRQs}$  \\
  \hline
   $\alpha_{min}$  & 0.167   & 0.208  \\
  $\alpha_{max}$ & 0.344   & 0.303  \\
  $A_{min}$ & 0.559  & 0.426\\
  $A_{max}$ &  0.143 & 0.204 \\
  $\tau_{min}$ & - & 0.282 \\
  $\tau_{max}$ & - & 0.246 \\
  Optimal & 0.281 & 0.277 \\
\hline
\end{tabular}
\end{table}

\section{Discussion}\label{discussion}

We have made a point of optimizing different models for BL Lacs and
FSRQs. An evaluation of whether this is indeed a necessary distinction
can be made after the fact, by comparing the optimal models between BL
Lacs and FSRQs. We find that FSRQs are: 
\begin{itemize}
\item Faster: BL Lacs statistically have lower Lorentz factors than
  FSRQs, as their $\Gamma-$distribution is steeper (Fig. \ref{plt_lorentz_joint}). 
\item Evolving: we have found that the luminosity distribution of
  FSRQs moves to higher luminosities with increasing redshift;
  conversely,  the  BL Lac redshift distribution with a luminosity function that evolves
  with redshift. In principle, this result could be affected by the
  redshift incompleteness of the BL Lac sample, which is larger than
  that of the FSRQ sample, if lower redshifts are preferentially
  easier to measure. However, there is no evidence for any evolution {\em even
    among lower redshifts} (for example from $z=0$ to $z=0.4$, where
  one would not expect to have our ability to measure redshifts to
  vary dramatically). If such a bias exists, it would have to be
  rather fine-tuned to exactly match the brightening of
  higher-redshift sources. This finding is consistent with gamma-ray
  studies of the BL Lac luminosity function \citep{Ajello2014}.  
\item Brighter:  Even though today BL Lacs are brighter than FSRQs
  (Fig.\ref{log_log_beamed_lum}), the situation is reversed at higher
  redshifts, due to the evolution of FSRQs. 
\end{itemize}
It is therefore particularly interesting that, despite their intrinsic
differences, these two classes appear
to have very consistent relativistic beaming ($\Gamma \theta$ distribution serves as a proxy) and timescale compression
distributions in flux-limited samples. Their intrinsic differences are
thus expected to be imprinted in their statistical properties in the
time domain measured in the observer frame. 

Throughout this work we have assumed a single value for the spectral
index ($s$) for all sources in each sample. In reality it is different
for every source and can evolve with time
\citep{Angelakis2012,Hovatta2014}. However, since this simple assumption has
yielded adequate results, we have not treated separately a spectral
index distribution, although such an extension could be added to our
model in a straight-forward fashion.

Single power law distributions for the Lorentz factors and the
luminosity function are sufficient to describe both samples: the K-S
test values for the FSRQs are 49.3\% for the apparent velocity and
8.4\% for the redshift distributions while for the BL Lacs 93.4\% for
the apparent velocity and 54.1\% for the redshift distributions. 

We have calculated, and we quote, mean values of the timescale
modulation factor, for a 1.5 Jy flux-limited sample. Since we have
shown that this distribution can be well-described by an exponential,
the mean is the only parameter needed to completely define
it. However, we emphasize that the value of the mean is a quantity
dependent on the flux limit. Applications to different samples need
to properly calculate the distribution appropriate to the relevant
flux limit. 

In addition we caution the reader that results of our models for lower flux limits involve extrapolation and should thus be treated with caution.

In this work, we have treated BL Lacs and FSRQs separately. The small number of BL Lacs make the BL Lac sample and the associated results more sensitive to contamination borderline shifting classification sources. However, it is striking that even with such a small sample, no agreement with even a mildly evolving luminosity function was possible.

The model optimized here was based on 15 GHz radio data. Any
application to the statistical interpretation of data obtained in
other frequencies should be done with two possible
caveats in mind. First, it is not necessarily obvious that a single
Lorentz factor can characterize the jet at all frequencies, or even at
all locations at the same frequency  (e.g.,
\citealp{Markos2004}). Second, 
a flux-limited sample at a different frequency range (e.g., gamma-ray
or optical), does not translate directly to a flux-limited sample in
radio (e.g., \citealp{Pavlidouff}). The scatter in the correlation between fluxes at different
frequencies must therefore be accounted for.

\section*{Acknowledgments}
We would like to thank Matt Lister, Talvikki Hovatta, and the anonymous referee for comments that helped improve this work.

This research was supported by the ``Aristeia'' Action of the  ``Operational Program Education and Lifelong Learning'' and is co-funded by the European Social Fund (ESF) and Greek National Resources, and by the European Commission Seventh Framework Program (FP7) through grants PCIG10-GA-2011-304001 ``JetPop'' and PIRSES-GA-2012-31578 ``EuroCal''. This research has made use of data from the MOJAVE database that is maintained by the MOJAVE team \citep{Lister2009-2}

\bibliographystyle{mn}
\bibliography{bibliography} 

\label{lastpage}

\end{document}